\documentclass{aa}
\usepackage{graphicx}
\usepackage{adjustbox}
\usepackage{natbib}
\usepackage{enumerate}
\usepackage{enumitem}
\usepackage{float}
\usepackage{siunitx}
\usepackage{txfonts}
\begin{document}

%
   \title{Searching for g modes: Part I. A new calibration of the GOLF instrument}

   \author{T.~Appourchaux
          \inst{1}
          \and
          P.~Boumier\inst{1}
          \and
          J.~W.~Leibacher\inst{1,2,3}
          \and
          T.~Corbard\inst{4}
          }

   \institute{Univ. Paris-Sud, Institut d'Astrophysique Spatiale, UMR 8617, CNRS, B\^atiment 121, 91405 Orsay Cedex, France
   \and
   National Solar Observatory, Boulder, CO, USA
   \and
   Lunar and Planetary Laboratory, University of Arizona, Tucson, AZ, USA
   \and
   Universit\'e C\^ote d'Azur, Observatoire de la C\^ote d'Azur, CNRS, Laboratoire Lagrange, CS 34229, Nice Cedex 4, France
          }

   \date{}

 
  \abstract
  {The recent claims of g-mode detection have restarted the search for these potentially extremely important modes.  These claims can be reassessed in view of the different data sets available from the SoHO instruments and ground-based instruments.}
   {We produce a new calibration of the GOLF data with a more consistent p-mode amplitude and a more consistent time shift correction compared to the time series used in the past.}
   {The calibration of 22 years of GOLF data is done with a simpler approach that uses only the predictive radial velocity of the SoHO spacecraft as a reference.  Using p modes, we measure and correct the time shift between ground- and space-based instruments and the GOLF instrument.}
   {The p-mode velocity calibration is now consistent to within a few percent with other instruments.  The remaining time shifts are {within $\pm$ 5 s for 99.8\% of the time series}.}
   {}

   \maketitle
%
\section{Introduction}
The detection of g modes remains a major objective of helioseismology.  The benefit of detecting these modes would be to obtain the structure and dynamics of the very inner
core of the Sun.  There have been several claims of g-mode detection \citep[See][for a review]{Appourchaux2010}.  Recently, \citet{Fossat2017} using the propagation time of the p-mode wave packet claimed to have detected the signature of g modes.  In order to test that detection claim, we made longer data sets using a new calibration strategy for the GOLF data.

Since the beginning of helioseismology, solar radial velocities have always been measured using solar spectral lines.  The intensities are typically measured in the blue and red wings ($I_b$ and $I_r$) of the line and the displacement is deduced by calibrating the ratio $\frac{I_r-I_b}{I_r+I_b}$ with respect to known radial velocities \citep[See][and references therein]{Elsworth1995}.  The purpose of the ratio is mainly to remove the effect of the Earth's atmospheric variations.  Concerning space-based instruments, the ratio is used for reducing the effect of the change of transmission in the course of the lifetime of the instrument.  For ground-based instruments, the ratio is measured at a very high cadence (from tens of Hz to a few kHz) while for space-based instruments, a slow cadence can be used (slower than 1 Hz).  { The} signal inferred from the ratio is somewhat affected by signals not related to radial velocities, but due to the effect of radiative transfer across the line \citep{Ulrich2000}.  

The calibration of the GOLF data was rendered more complicated by the fact that the two-wing measurement could not be made after 31 March 1996 \citep[See][]{Gelly2002}.  This introduced an additional effect on the p-mode velocity signal since on either wing the intensities are intrinsically modulated by intensity fluctuations due to the p modes themselves and the granulation background.  The fraction of the intensity fluctuations to velocity fluctuations due to p modes is about 0.12 \citep{Renaud1999}.  The measurement of the GOLF instrument was then done using only one wing (red or blue) using three different methods \citep{Ulrich2000,Gelly2002,Garcia2005}.  \citet{Ulrich2000} relied on a detailed modelling of the line profile for inferring the residual velocities.  \citet{Gelly2002} tried to minimise the yearly modulation of p-mode power to optimise the calibration. The change of p-mode power from 1996 to 2002 due to
solar activity was 10\%, typical of what is observed by the Birmingham Solar Oscillation Network\footnote{BiSON, See \citet{WJC1996}} \citep[See][]{Howe2015}.  \citet{Garcia2005} relied on instrumental calibrations and a non-linear calibration method developed by \cite{Palle1993}.  

\begin{figure*}[!]
\centerline{
\vbox{
\hbox{
\includegraphics[width=6 cm,angle=90]{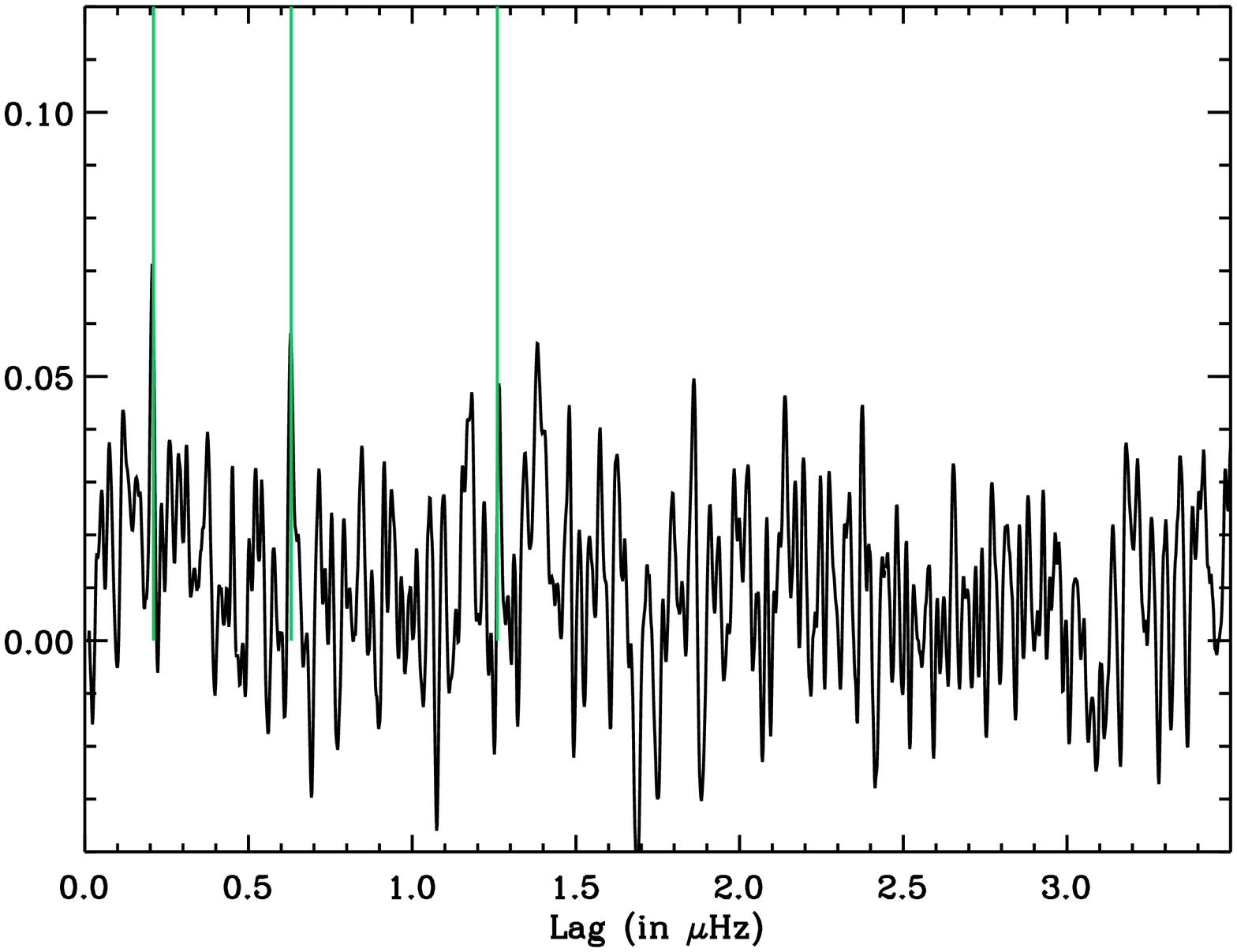}
\includegraphics[width=6 cm,angle=90]{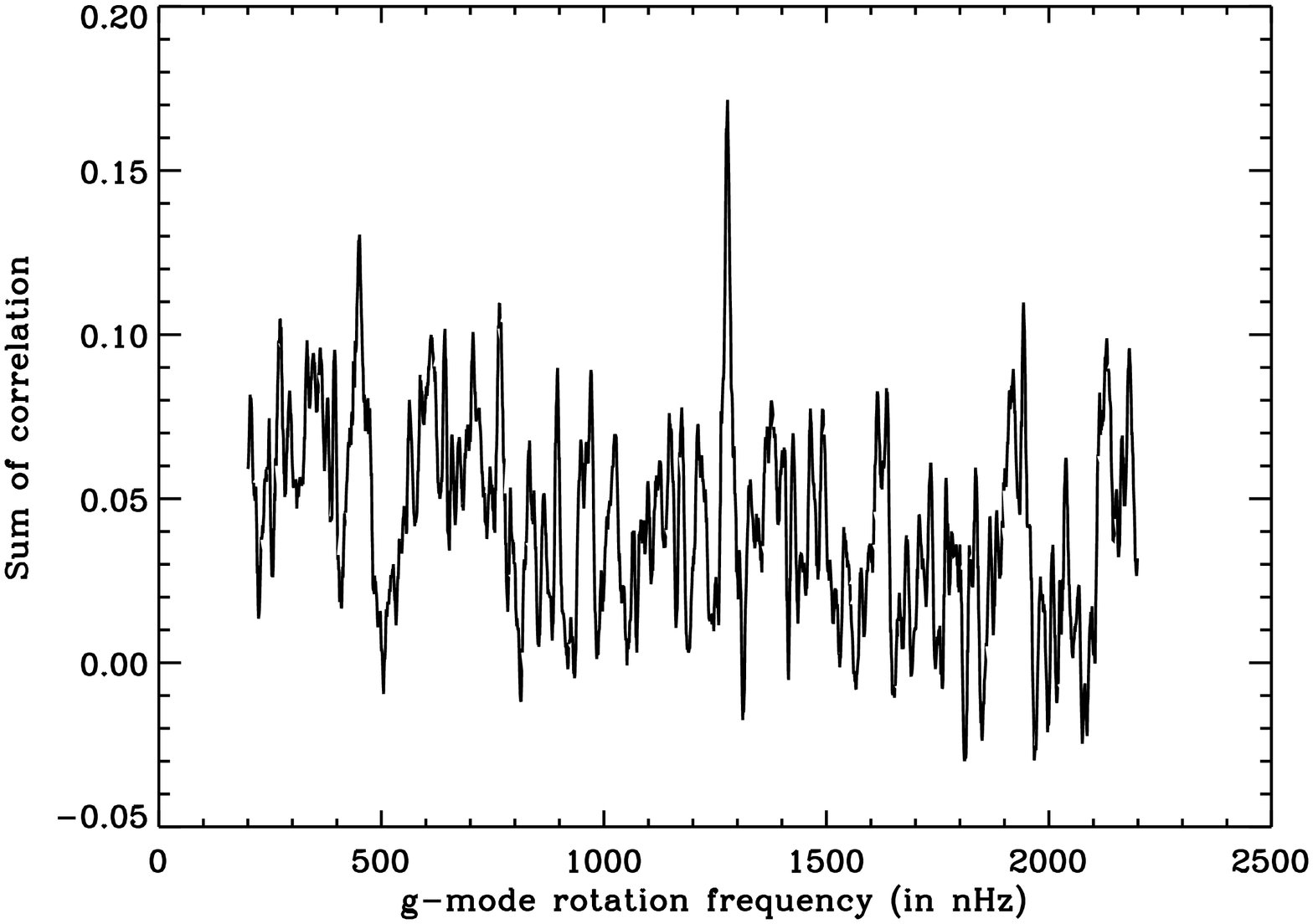}}
\hbox{
\includegraphics[width=6 cm,angle=90]{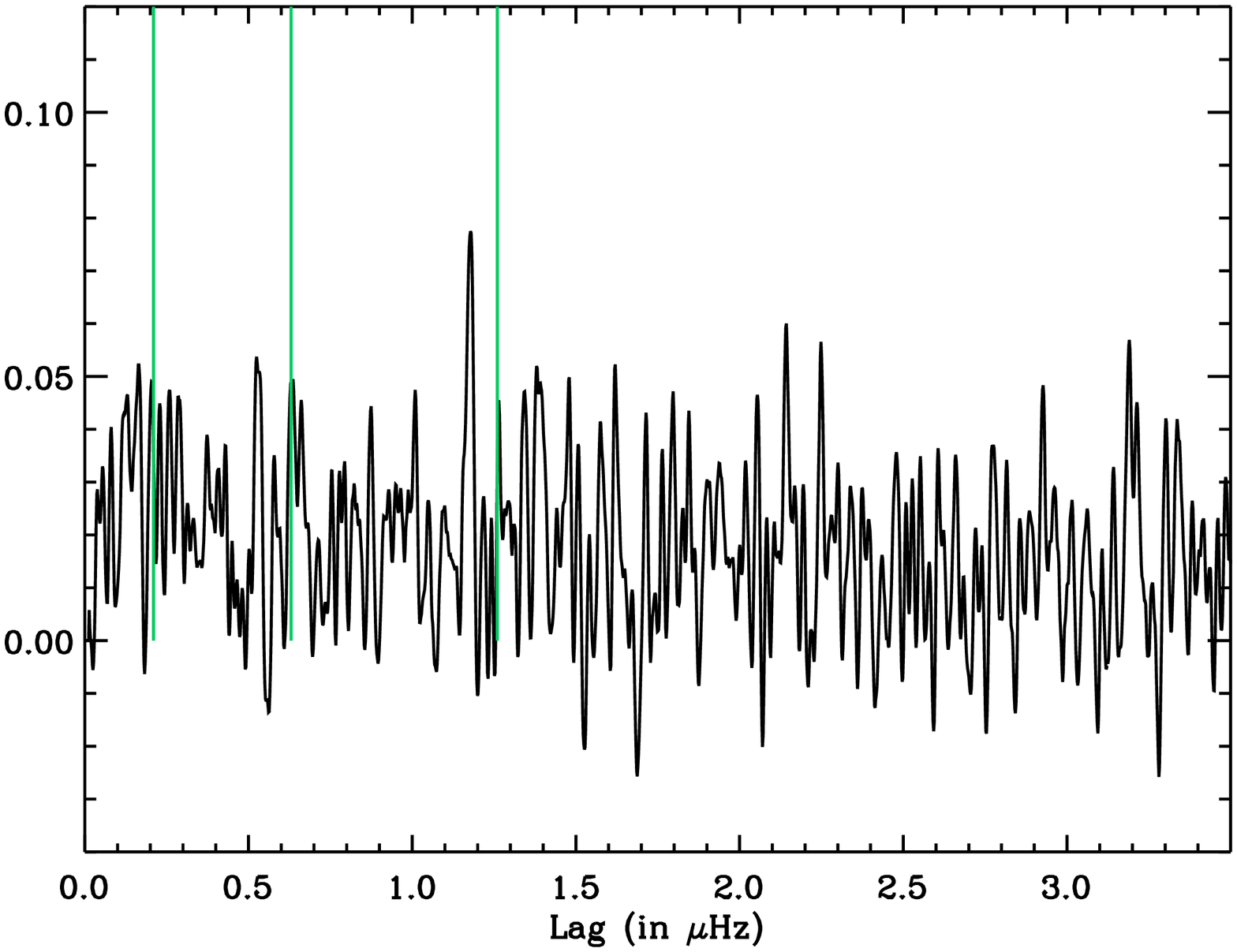}
\includegraphics[width=6 cm,angle=90]{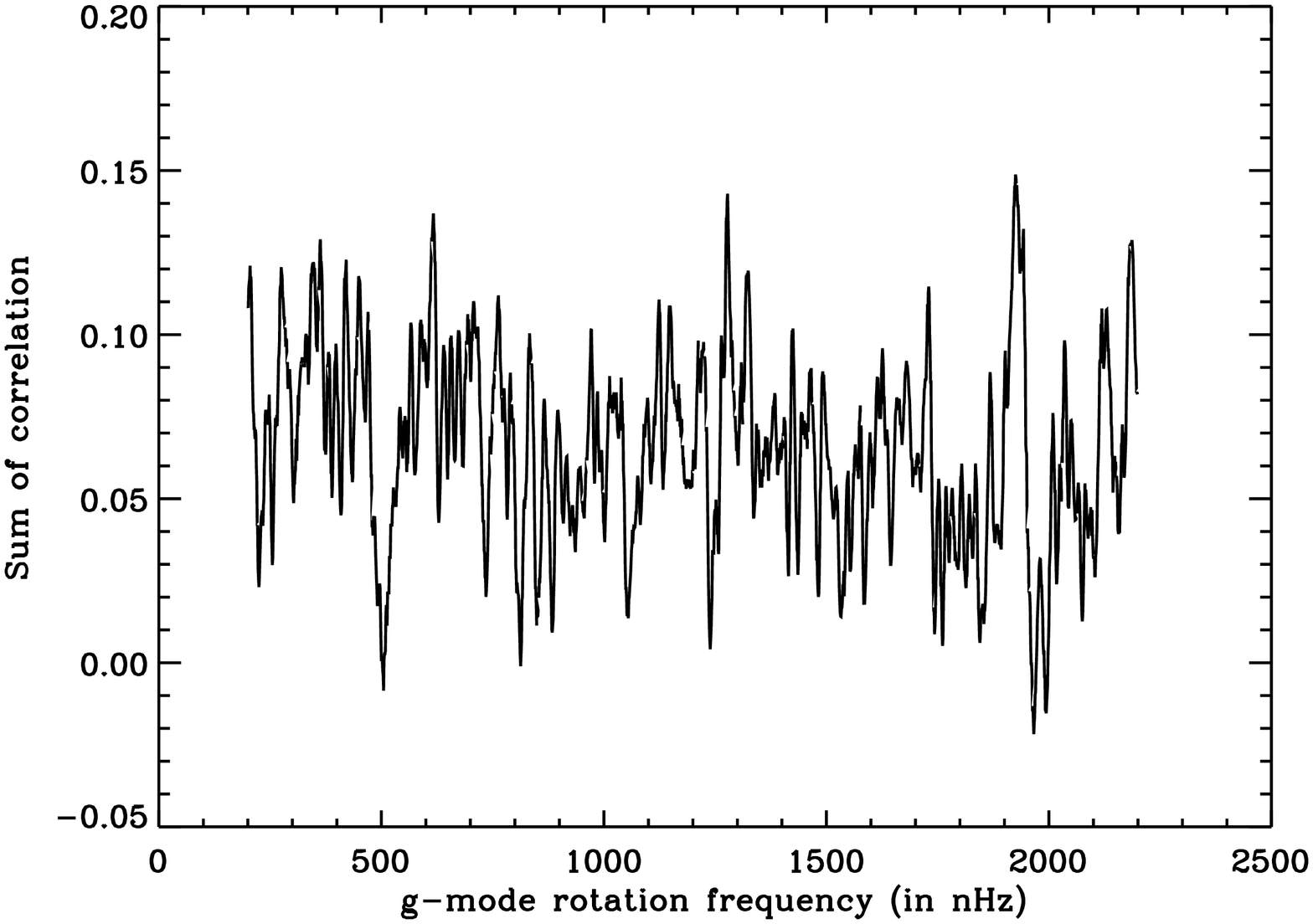}}
}}
\caption{(Left) Correlation of the power spectrum as obtained by \citet{Fossat2017} as a function of frequency lag for two different time series sampled at 80 s (top) and 60 s (bottom); this is comparable to Fig.~10 of \citet{Fossat2017}.  The green vertical lines correspond to frequencies at 210 nHz, 630 nHz, and 1260 nHz.  (Right) Sum of the correlation for $l=1$ and $l=2$ modes as obtained by \citet{Fossat2017} as a function of rotation frequency for two different time series sampled at 80 s (top) and 60 s (bottom); this is comparable to Fig.~16 of \citet{Fossat2017}}
\label{Fossat}
\end{figure*}

The recent g-mode detection claim of \citet{Fossat2017} was performed using the calibration done by  \citet{Garcia2005}.  There are two types of data available on the GOLF website\footnote{See www.ias.u-psud.fr/golf/templates/access.html}: those used by 
\citet{Fossat2017} sampled at 80 s, and a time series sampled at 60 s.  Both data have the same length and are produced using{ } the same time series sampled at 20 s but binning over 4 and 3 samples,
respectively (Garc\'\i a, private communication, 2018).  In the course of reproducing the findings of \citet{Fossat2017}, we used both time series to produce our own version of their Figs.~10 and 16.  Figure~\ref{Fossat} shows the output of the procedure used by \citet{Fossat2017} on two different times series.  It is clear that Figs.~10 and 16 of \citet{Fossat2017} cannot be reproduced with the time series sampled at 60 s.  Very recently, \citet{Schunker2018} reproduced Fig.~10 of \citet{Fossat2017}.  They showed that using different fitting procedures, the prominent peaks at 210 nHz and its acolytes would smear out or even disappear, as seen for the time series sampled at 60 s.  This leads us to investigate how the GOLF data are calibrated and whether a different calibration might or might not reproduce the results of \citet{Fossat2017}.

This work has been divided in two parts.  The present paper (Part I) explains how a new calibration of the GOLF data has been obtained.  The accompanying paper (Part II) investigates the results obtained by \citet{Fossat2017} and compares it with different time series.  The first section of this present paper explains the calibration procedure, the extraction of the velocity, and the time corrections applied.  The second section compares the calibration results obtained with other data sets.  Subsequently, we present our conclusions.



\section{A new calibration of the GOLF data}
\subsection{Extraction of the velocity}
\begin{figure}[!]
\centerline{
\includegraphics[width=6 cm,angle=90]{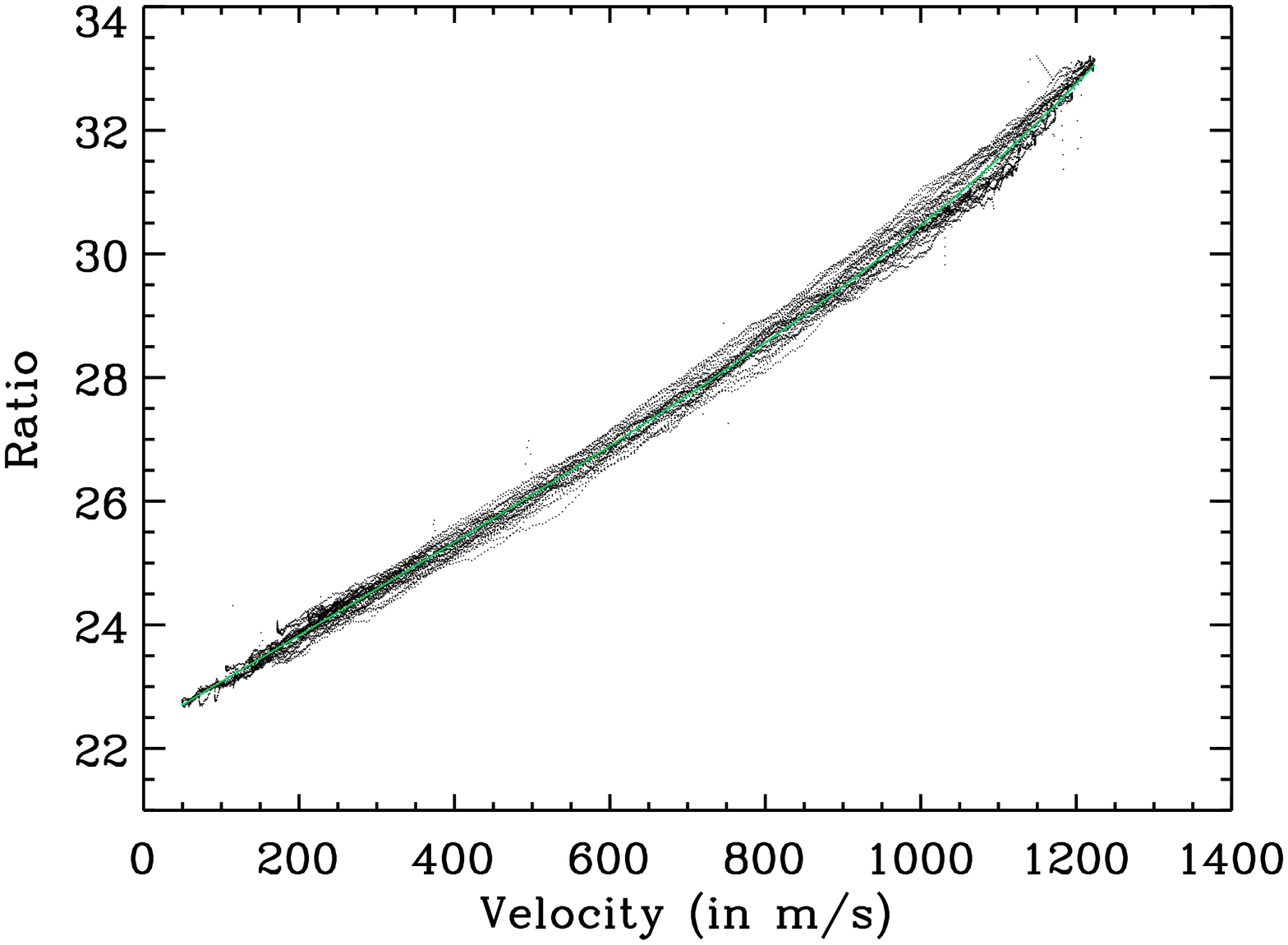}}
\hfill
\caption{$X$ ratio as a function of the velocity for PM1: one point per day.  The fit of the $X$ ratio is in green.  The fit is for the blue wing after February 2002.}
\label{fit_X}
\end{figure}

\begin{figure}[htbp]
\centering
\includegraphics[width=6 cm,angle=90]{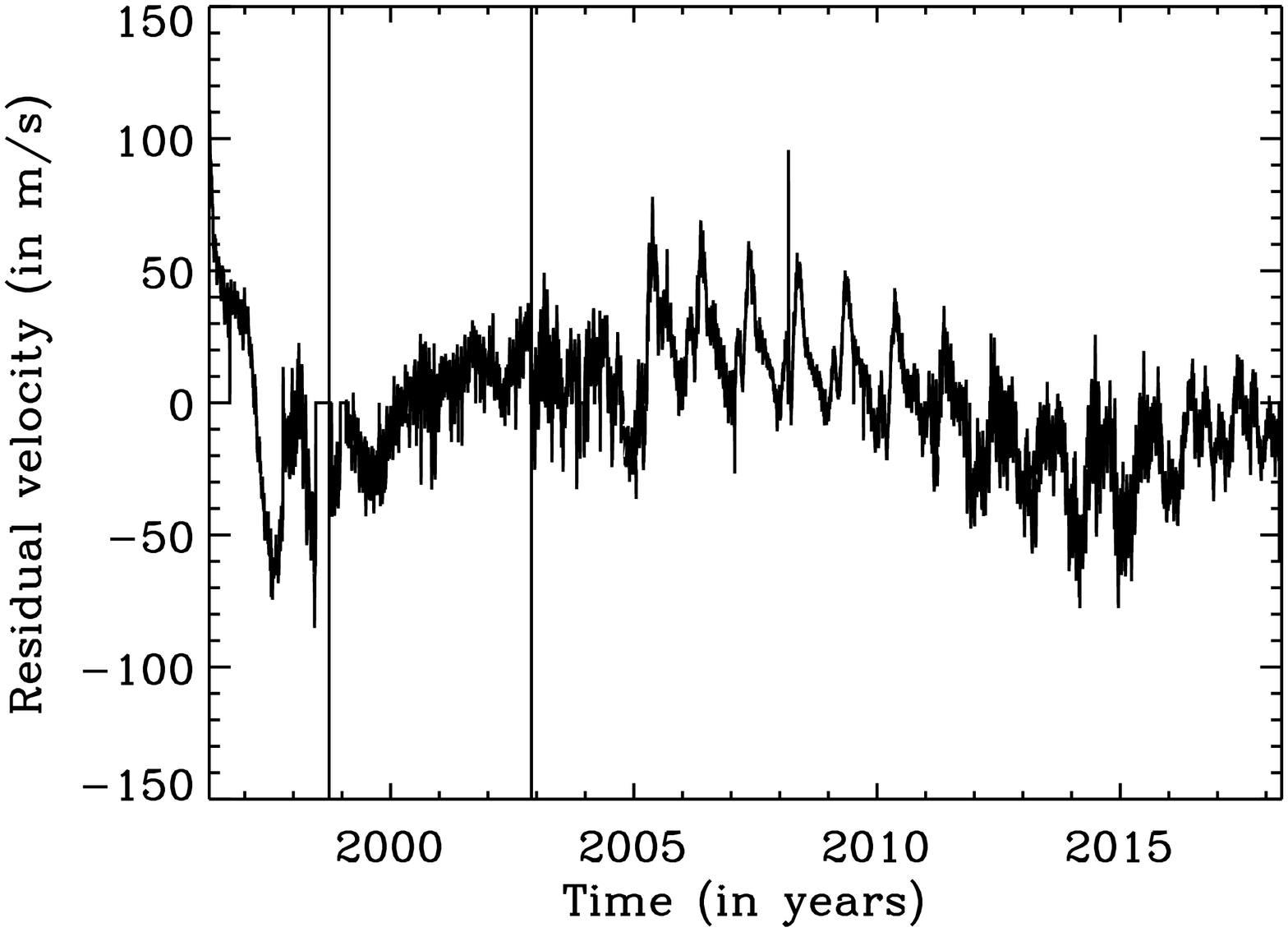}
\caption{Velocity residuals for PM1 as a function of time: one point per day.  The two vertical lines indicate when the modes of operation changed from blue wing to red wing and vice versa.}
\label{residual}
\end{figure}

The data used for the new calibration start on 11 April 1996 and end on 10 April 2018.  The technique used for the calibration is derived from
that of \citet{Garcia2005} using the so-called $X$ method.  Since GOLF operates only in one wing of the Sodium lines, the following ratio is computed
as a proxy to the two-wing signal:
\begin{equation}
X=\frac{1}{2} \frac{P^{+}+P^{-}}{<P^{+}-P^{-}>}
\label{X}
,\end{equation}
where $P^{+}$ and $P^{-}$ are the signals measured in the same wing (blue or red) using a weak magnetic field modulation that induces
a wavelength change.  The bracket denotes a low-frequency filtering of the
slope proxy (i.e. the derivative).   The division by $<P^{+}-P^{-}>$ is needed since it provides a way to normalise to the slope of the line.  The procedure for computing the radial velocity starts by computing the photomultiplier (PM) signals as follows:
\begin{enumerate}
\item Read daily data for both PMs and magnetic modulation \citep[4 signals as $P^{+}_{1,2}$, $P^{-}_{1,2}$; see][for details]{Gabriel95}.
\item Compute daily median of the slope proxy ($P^{+}_{1,2}-P^{-}_{1,2}$) for low frequency filtering.
\item { Either no time correction is performed or if a time correction is measured,} shift daily data by integer multiples of 10\,s (No interpolation is used to avoid introducing additional noise).
\item Concatenate all of the daily data and save the the four signals and the slope proxy.
\end{enumerate}
We note that { the time correction in} Step 3 is { only} made { in a second iteration} after the time shift measurement { is performed on the unshifted time series}.  { The procedure for the time shift measurement} is explained in Sect.  2.2.  In a second stage, we compute the $X_{1,2}$ ratio as follows:
\begin{enumerate}
\item Bin the data originally sampled at 10 s over two samples resulting in a 20-s cadence.
\item Using a spline, interpolate the daily slope proxy over the 20-s sample to a uniform temporal sampling.
\item Compute the { $X_{1}$ and  $X_{2}$ ratios} according to Eq.\,(\ref{X}).
\item Cut the { time series of the $X_{1}$ and  $X_{2}$ ratios} in sub-series of 20 days.
\item Interpolate to remove one-sample spikes greater than 11\% of the median of the sub-series.
\item Compute residuals of the sub-series with respect to a two-day smoothed version of the sub-series.
\item Detect variations greater than 1\% in the residuals then replace the remaining outliers by interpolation and remove the slow changes using a sixth-order polynomial.
\end{enumerate}
The cut-offs used for filtering the outliers are a compromise between the level of the outliers and the signals to be kept; this could be perceived as arbitrary but in fact this procedure was chosen according to our experience with the data.  The calibration of the { $X_{1}$ and  $X_{2}$} ratios with respect to the solar radial velocity is done using the technique described in \citet{Garcia2005}, in which they assume that the ratio $X$ can be modelled as follows:
\begin{equation}
X=f(V)=\frac{a+bV}{1+\alpha V^2+\beta V^4}
\label{Palle}
,\end{equation}
where $a, b, \alpha$ and $\beta$ are parameters, and $V=V_{\rm grav}+V_{\rm orb}$ is the radial velocity including the gravitational redshift ($V_{\rm grav}$=636~m\,s$^{-1}$) and the contribution generated by the spacecraft orbiting the Sun ($V_{\rm orb}$).  The residual velocity ($V_{X}$, made of the oscillation signals and other contributions) is excluded from the model since this is what we want to extract.  The orbital spacecraft velocities are obtained from the VIRGO Data Center, which provides computation of the solar radial velocity on a 60-s cadence.  The velocity used here is the predictive, not the measured, velocity since the SoHO operation team stopped providing the reconstructed velocity after 1998 because the predictive one was good enough.  The spacecraft velocity is then interpolated onto the sample series (20-s cadence).  In theory, one would invert Eq.~(\ref{Palle}) to obtain the residual velocity ($V_{X}$) as $V_X=f^{-1}(X)-V$ but this is not done in practice.  The residual velocity is inferred as in \citet{Garcia2005} using the following computation of the residuals:
\begin{equation}
V_{X}=\frac{1}{b}\left(X\,(1+\alpha V^2+\beta V^4)-a\right)-V
\label{Palle1}
.\end{equation}
This so-called inversion performs better than any other true inversion obtained by solving Eq.~(\ref{Palle}).  The true inversion provides yearly modulations of the p-mode amplitude that are
not physical.  We also tried using the calibration procedure of \citet{Elsworth1995} consisting in fitting a polynomial to the ratio $X$ as a function of $V$, then using the derivative to obtain the velocity residuals, but this scheme also produces the same yearly modulation.  The reason for the exceptional performance of Eqs.~(\ref{Palle}) and  (\ref{Palle1}) is not yet fully understood.  The calibration of the velocity is done for three time segments, each corresponding to a different operational mode of GOLF, as in \citet{Garcia2005}.  The ratio is fitted in two passes; on the second pass, residuals larger than 150 m\,s$^{-1}$ are excluded from the fit (about 0.2\% of the points are excluded).  A fitted ratio is shown for the blue wing in Fig. \ref{fit_X} for PM1.  We compute the residual using Eq.~(\ref{Palle1}).  Then on that residual, we detect spikes greater than 12 m\,s$^{-1}$ by applying a high pass filter based on a two-day triangular smoothing; about 0.15\% of the points are rejected.  Table~\ref{fitted} gives the fitted parameters { according to Eq.\,(\ref{Palle}) for the ratios $X_{1}$ and $X_{2}$} .  The final velocity is the average of the velocity residuals of the times series of PM1 and PM2.  Figure \ref{residual} gives the final residual velocity for PM1.  There is still a yearly modulation of the residual velocity that is related to the variation of the temperature along the orbit which is not properly taken into account.  Nevertheless this residual modulation has no impact on the p-mode amplitude (See Sect. 3).


\begin{table*}[t]
\caption{Fitted parameters of Eq.~(\ref{Palle}) with their error bars.  The first column gives the PM identification and in which wing the measurement was made.  The other columns give the parameters of Eq.~(\ref{Palle}) together with their formal error bars.}             
\label{fitted}      
\centering                          
\begin{tabular}{c 
cccccccc}        
\hline                 
\hline     
Wing&{$a$} &{$\sigma_{a}$}&{$b$}&{$\sigma_{b}$}&{$\alpha$} &{$\sigma_{\alpha}$}&{$\beta$} &{$\sigma_{\beta}$}\\
&&&10$^{-3}$\,\si{s.m^{-1}}&10$^{-3}$\,\si{s.m^{-1}}&10$^{-7}$\,\si{s^{2}.m^{-2}}&10$^{-7}$\,\si{s^{2}.m^{-2}}&10$^{-14}$\,\si{s^{4}.m^{-4}}&{10$^{-14}$\,\si{s^{4}.m^{-4}}}\\
\hline  
\hline                           
PM1 Blue&23.5720&0.0018&+6.900&0.0092&--0.4761&0.0045 &--0.0015 &0.0228\\
PM2 Blue&23.7150&0.0018&+7.255&0.0093&--0.2771&0.0045&--0.8282 &0.0224\\
PM1 Red&22.1086&0.0013&--5.970&0.0075&--1.0538&0.0057&+0.1856&0.0235\\
PM2 Red&22.0714&0.0013&--6.100&0.0075&--1.0929&0.0057&+0.0540&0.0235\\
New PM1 Blue&22.3316&0.0008&+7.472&0.0036&--0.0069&0.0016&--2.0874&0.0068\\
New PM2 Blue&22.1302&0.0008&+7.521&0.0036&--0.0344&0.0016&--1.7253&0.0068\\
\hline  
\hline  
\end{tabular}
\end{table*}

\subsection{Correction of time shifts}
As outlined in \citet{Garcia2005}, the GOLF data may lose synchronisation with respect to the basic temporal cadence.  In principle, ancillary data allow us to quantify the resulting induced shifts thanks to a daily pulse (DP) generated on board SoHO. The precise time datation according to the Temps Atomique International (TAI) is available in the header of the raw daily data files. However, in a few cases, the GOLF clock could not be synchronised on the DP, and in order to check for potential unknown time jumps, we used data from other instruments such as the Global Oscillation Network Group\footnote{GONG, See \citet{Harvey1996}}, the Birmingham Solar Oscillation Network,\footnote{BiSON, See \citet{WJC1996}} and the Sun Photometer Monitor of the Variability of Irradiance and Gravity Oscillations instrument\footnote{VIRGO / SPM, See \citet{CF97}}.  The measurement of the time shift is done using the p modes themselves with the following procedure:
\begin{enumerate}
\item Select a sub-series from two instruments (e.g. GOLF, BiSON, GONG, SPM) with a typical duration of one day.  When the cadence differs, resample the sub-series to a common temporal cadence of 60 s using linear interpolation (e.g. for BiSON).  The integration is one day as well as the sampling (for display purposes, other integration times of 10 days and 45 days have been used with one-day sampling).
\item Compute the backward difference filter (BDF)\footnote{See Appendix A for explanation}, for each instrument but GONG, for which it has already been applied.
\item Compute the cross correlation $C$ for a range of $\pm$ 13 mins by steps of 1 min.
\item Compute the cross correlation envelope $C_{\rm env}$ using the Hilbert transform $\cal{H}$ ($C_{\rm env}=\sqrt{C^2+{\cal{H}}(C)^2}$).\footnote{for a sinusoidal function $C$, $C_{\rm env}$ provides its amplitude \citep{Feldman}}
\item Fit the cross correlation envelope using least squares with a Gaussian to obtain an estimate of the time shift between the two data sets.  This
estimation is used for the following step.
\item Fit the cross correlation using least squares with a time-shifted cosine function modulated by a Gaussian envelope.  The Gaussian envelope and the cosine function have the same time shift.
\end{enumerate}
The BDF is used to reduce the low-frequency solar noise especially when correlations are made between intensity and velocity signals.  Since GONG uses the BDF by default\footnote{See gong.nso.edu/data/pipe\_stages/GONG\_DowNStream\_Pipeline.html}, this filter is not applied to GONG.  The precision of the time shift obtained by fitting the correlation of the p modes is far more precise than the fitting of the correlation provided by the envelope.  The former is related to the p-mode phase velocity while the latter is related to the p-mode group velocity.  

\begin{figure}[htbp]
\centering
\hbox{
\hspace{0cm}
\includegraphics[width=6 cm,angle=90]{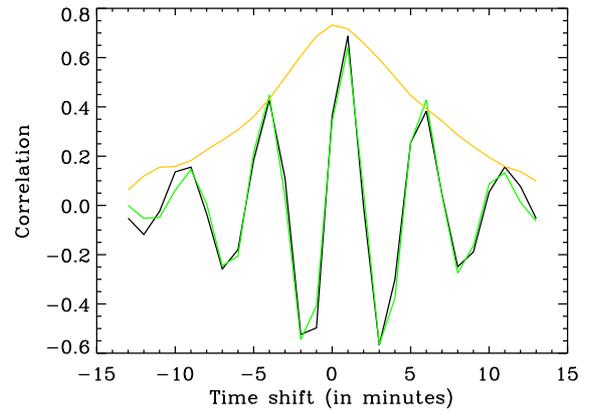}
}
\caption{Cross correlation between the first difference of the GOLF radial velocity and the first difference of the $l=0$ GONG radial velocity as a function of lag in minutes (black line), with the envelope shown (orange line) and the fitted cross correlation (green line).}
\label{time_shift_fit}
\end{figure}

\begin{figure}[htbp]
\centering
\hbox{
\includegraphics[width=6 cm,angle=90]{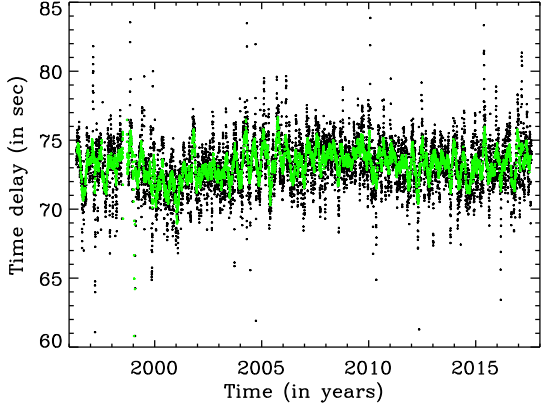}
}
\caption{Time difference between the SPM Blue / VIRGO data and the $l=0$ GONG data as a function of time computed using a 10-day window (black dots) or a 45-day window (green dots).}
\label{delay_SPM_GONG}
\end{figure}

Figure~\ref{time_shift_fit} gives an example of what is fitted.  We found that the GONG data for $l=0$ and the VIRGO / SPM data do not present any temporal jumps as shown in Fig.~\ref{delay_SPM_GONG}.  As it is unlikely that both time series would have time jumps at the same date, the blue SPM and GONG then provide two references for a constant time base.   The GONG data were then used as a time reference to check and correct GOLF datation.  { On close inspection, Fig.~\ref{delay_SPM_GONG} shows that the time delay in this time series is modulated with a periodicity of almost 6 months.  This is due to the SoHO halo orbit\footnote{See www.esa.int/esapub/bulletin/bullet88/images/vand88f3.gif}, which has a periodicity of 178 days resulting in a time modulation of 1.54 s.  The periodicity of 178 days is primarily constrained by the solar radiation pressure \citep{Halo2016}.}

\begin{figure}[htbp]
\centering
\vbox{
\includegraphics[width=6 cm,angle=90]{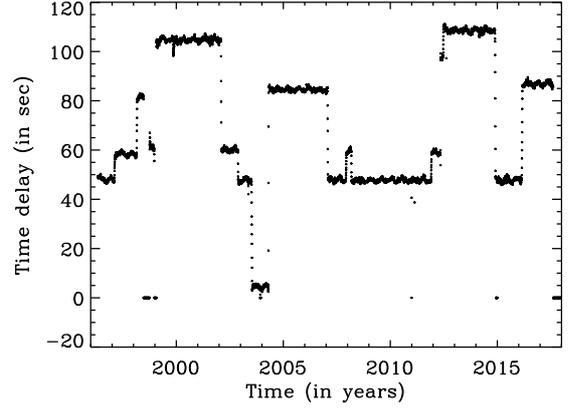}
\includegraphics[width=6 cm,angle=90]{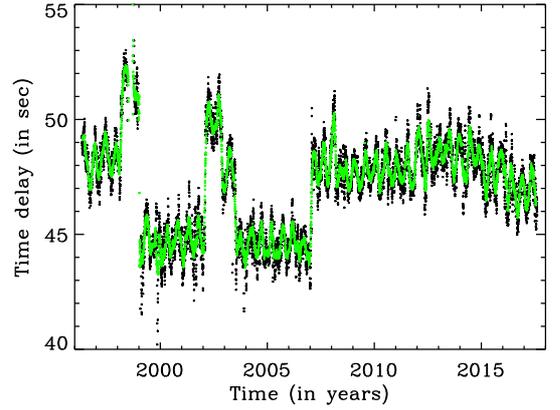}
}
\caption{Time difference between the newly calibrated GOLF data and the $l=0$ GONG data as a function of time; (top) Unshifted GOLF data; (bottom) Shifted GOLF data according to Table \ref{time_shift} computed using a 10-day integration window (black dots) or a 45-day integration window (green dots).  The median value of the time shift for the corrected time series is { 47.3} s. { About 99.8 \% of the} corrected values are within 5 s of this median value.}
\label{delay_GOLF_GONG}
\end{figure}

Figure \ref{delay_GOLF_GONG} shows the measured time delay of the unshifted GOLF time series together with the time delay of the corrected time series using the shifts listed in Table~\ref{time_shift}.   The measured time shifts as given in Table~\ref{time_shift} are used for correcting the time series after the listed dates of the table.  We used an integration time of one day to find the exact dates of the jumps.  After correction, the time series are checked again for potential errors in the date of the jumps.  We note that the intrinsic time shift of about 12 s measured by \citet{Renaud1999} between the blue wing and the red wing is automatically corrected by our procedure.  We also outline that the TAI time retrieved from the header of the file does not always reflect the presence of a time jump, and even sometimes the time jump is detected before the exact date provided by the change deduced from the TAI time.  { As for SPM, we can see the six-month modulation in the corrected time series even more clearly in Fig. \ref{delay_GOLF_GONG}.}

\begin{figure}[htbp]
\centering
\vbox{
\includegraphics[width=6 cm,angle=90]{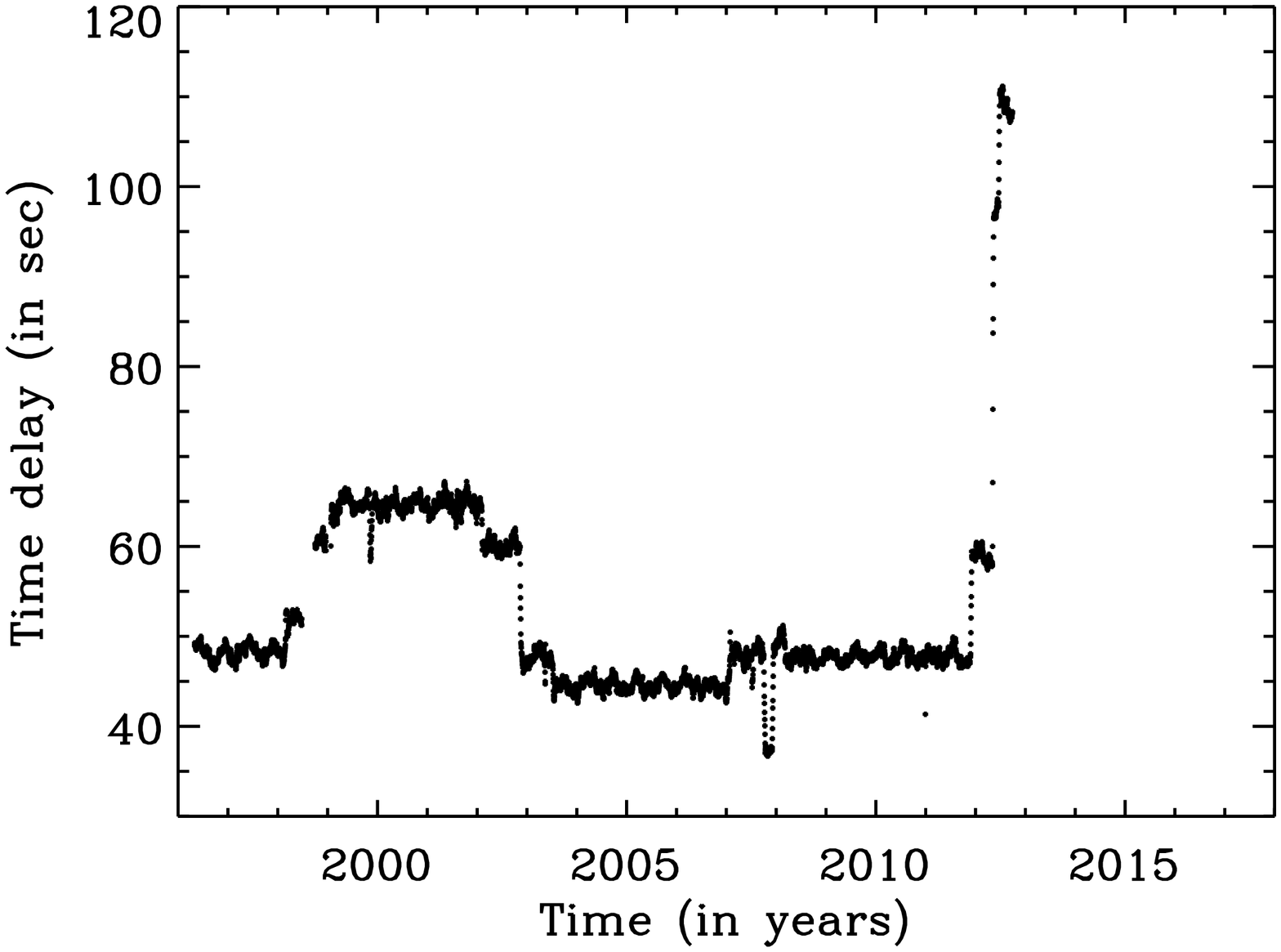}
\includegraphics[width=6 cm,angle=90]{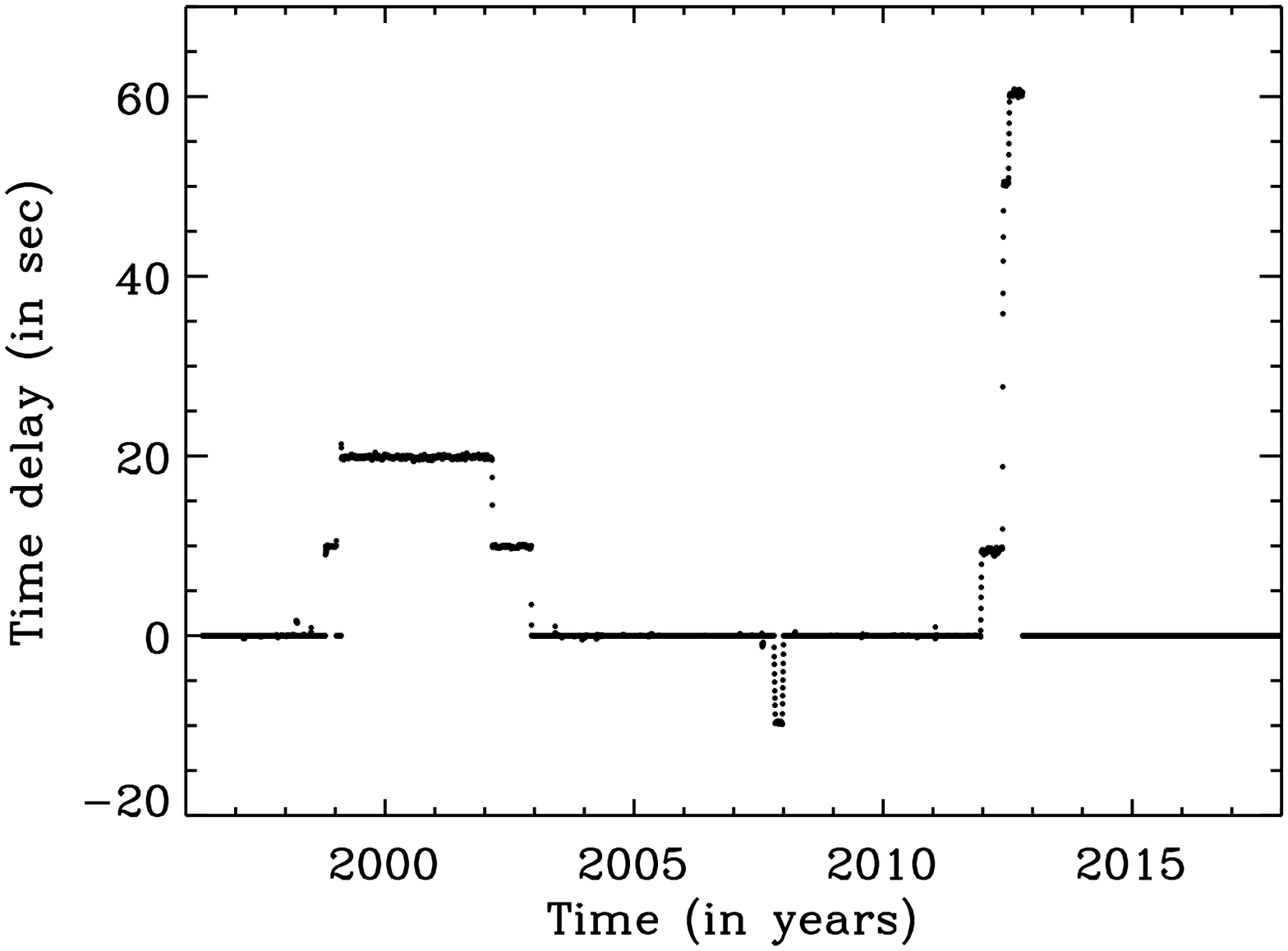}
}
\caption{(Top) Time difference between the GOLF time series sampled at 60 s and the GONG $l=0$ data as a function of time. (Bottom)  Time difference between the GOLF time series used by \citet{Fossat2017} and the the newly calibrated GOLF data as a function of time, both sampled at 80 s.}
\label{delay_GOLF_rafa_GONG}
\end{figure}

The typical rms precision obtained is of the order of 0.5 and 2.5 s for 10 days of integration, for GOLF and SPM, respectively.  Figure~\ref{delay_GOLF_rafa_GONG} shows the time delay of the time series used by \citet{Fossat2017} with respect to GONG and to the newly corrected GOLF time series.  There are seven time shifts not corrected in the time series used by \citet{Fossat2017}.  When the time shift agrees between the two versions of the GOLF time series, the resulting mean time difference is 0.4 $\pm$ 42 ms.


\begin{table*}[t]
\caption{Required time shift for correction of time jumps in GOLF.  The first column is the starting date of the shift, the second column is
the residual TAI in seconds of the first block of
each day, the third column is the measured time shift compared to GONG, the fourth column is the time shift with respect to the start of the time series (deduced from the third column), the fifth column is the residual TAI time after correction, and the sixth column is the residual time shift after correction. LOBT stands for Local On-Board Time.}             
\label{time_shift}      
\centering                          
\sisetup{
table-figures-integer = 2,
table-figures-decimal = 4
}
\begin{tabular}{
r 
S[table-number-alignment = right]
S[
separate-uncertainty,
table-figures-uncertainty = 1
]
S[table-number-alignment = center]
S[table-number-alignment = center]
S[table-number-alignment = center]
l}        
\hline                 
\hline     
{Date}&{TAI}&{$\tau_{\rm{meas.}}$}&{$\Delta\tau$}&{TAI delay}&{Final delay}&{Comments}\\
&{(s)}&{(s)}&{(s)}&{(s)}&{(s)}\\
\hline  
\hline                           
11 April 1996&30.0137&48.2 (8) &0.&0.&0.& Time reference for the whole series\\
16 February 1997&40.4536&58.3 (7)&10.1&0.44&0.1&LOBT change occurred on 25 February 1997\\
4 March 1998&63.3188&81.8 (7) &33.6&3.31&3.6&\\
9 October 1998&30.0083&61.3 (10)&13.1&-10.00&3.1&red-wing phase effect\\
2 February 1999&74.8770&104.6 (10)&56.4&-15.14&-3.6&LOBT change occurred on 3 February 1999\\
8 February 2002&30.0186&60.1 (7)&11.9&-10.00&1.9&red-wing phase effect\\
24 November 2002&{-}&47.8 (8)&-0.4&{-}&-0.4&No LOBT change.  Change of wing\\
15 July 2003&31.7793&4.7 (5)&-43.5&41.8&-3.5&LOBT not reliable\\
26 April 2004&66.6792&84.5 (7)&36.3&-3.33&-3.7&\\
27 January 2007&30.0239&47.9 (8)&-0.3&0.&-0.3&LOBT change occurred on 28 January 2007\\
10 December 2007&35.8530&59.6(8)&11.4&-4.16&1.4&\\
6 March 2008&30.0132&47.7 (7) &-0.5&0.&-0.5&LOBT change occurred on 7 March 2008\\
1 December 2011&{-}&59.0 (6)&10.8&{-}&0.8&No LOBT change\\
9 May 2012&79.2715&98.5 (11)&50.3&-0.74&0.3&LOBT change occurred on 12 May 2012\\
23 June 2012&{-}&108.8 (15)&60.6&{-}&0.6&No LOBT change\\
23 November 2014&18.5044&46.9 (14)&-1.3&-11.51&-1.3&LOBT change occurred on 24 November 2014\\
2 September 2015&26.3955&46.9 (14) &-1.3&-3.62&-1.3&LOBT change not detected by time correlation\\
26 February 2016&60.0088&86.8 (14)&38.6&-1.00&-1.4&LOBT change occurred on 27 February 2016\\
\hline  
\hline  
\end{tabular}
\end{table*}



\begin{figure}[htbp]
\centering
\vbox{
\includegraphics[width=6 cm,angle=90]{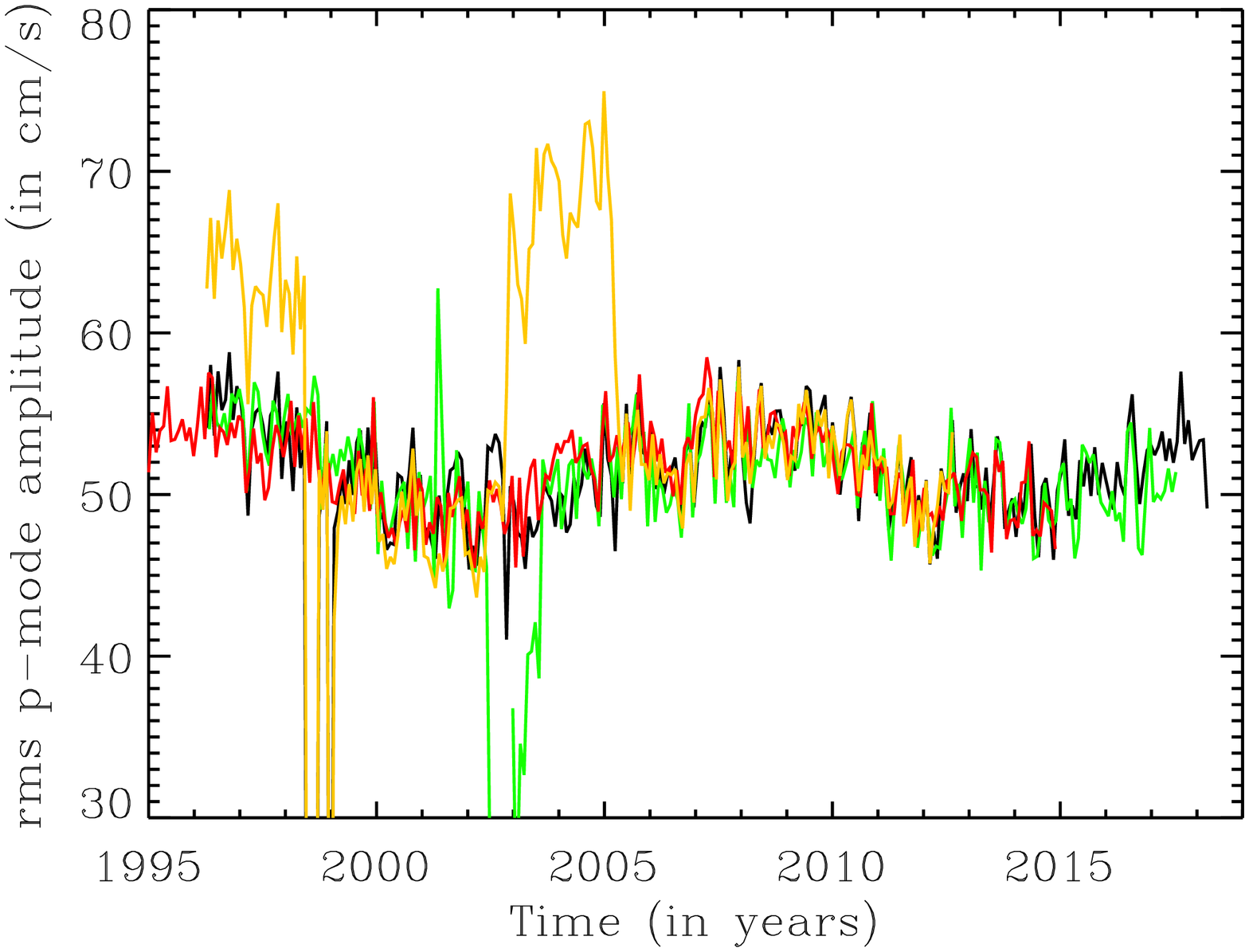}
\includegraphics[width=6 cm,angle=90]{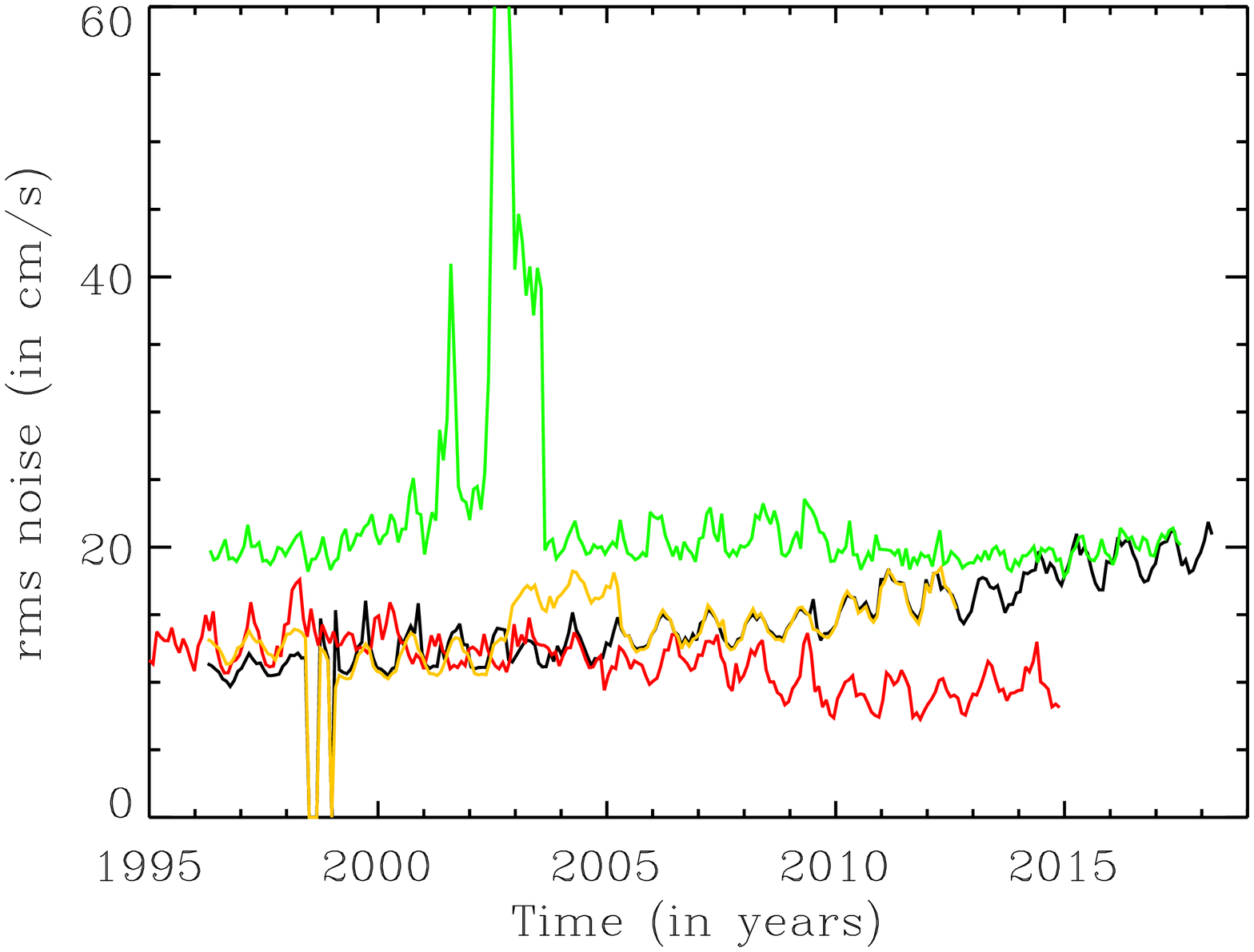}
}
\caption{(top) rms mode amplitude as a function of time for the newly calibrated GOLF data (black), for the GOLF data as used in \citet{Fossat2017} (orange), for the BiSON data (red), and for the GONG data (green). (bottom) rms noise as a function of time.  { The sub-series are 30 days in length}.}
\label{amplitude}
\end{figure}

\section{Comparison of the calibration with other data sets}
In order to double check the velocity calibration and time correction, we used data from GONG $l=0$ and from BiSON\footnote{Performance-check data available on bison.ph.bham.ac.uk/} and compared them with the previous calibration of the time series used by \citet{Fossat2017}.  Figures \ref{amplitude} shows the comparison of the rms p-mode amplitude and the rms p-mode noise for these different time series.  The computation of the rms amplitude and noise was done as follows.
\begin{enumerate}
\item { Select a sub-series of the data set, then compute the power spectrum.  Typical duration of the sub-series is 30 days.}
\item Compute the integrated power ($P_{\rm T}$) between 2500 $\mu$Hz and 3500 $\mu$Hz normalised by the integration factor of the sampling window (i.e. divide the power spectrum by $[{\rm sinc}(\pi \nu \tau)]^2$, where $\tau$ is the integration time of a sample).  This provides a proxy of the power of the sum of the noise power and the p-mode power.  
\item Compute the integrated power between 1000 $\mu$Hz and 1500 $\mu$Hz normalised by the integration factor of the sampling window.  For providing a proxy of the noise power in the p-mode range, we scaled this power by a factor taking account of the frequency bandwidth (factor 2) and extrapolation factor (factor 1/$\sqrt{2}$); we then multiply this power by $\sqrt{2}$  to obtain the proxy ($P_{\rm noise}$). 
\item The final p-mode power is obtained as  $P_{\rm p-mode}$=$P_{\rm T}-P_{\rm noise.}$
\item The rms amplitude is then $A_{\rm p-mode}=\sqrt{P_{\rm p-mode}}$, while the rms noise is $A_{\rm noise}=\sqrt{P_{\rm noise.}}$
\item Apply an ad-hoc formation height correction for the amplitudes (if required).
\end{enumerate}
We apply this procedure to all data but GONG.  For GONG, additional corrections of the power spectrum were required because of the BDF used, which requires us to divide by the filtering factor $T(\nu)=4 \sin^2(\pi \nu \Delta t)$, where $\Delta t$ is the sampling cadence.  For the p-mode frequency range and a 60-s cadence, the filtering factor in power ranges from 0.8 to 1.5, while for the noise it ranges from 0.14 to 0.31.    

For the last step, the formation height was corrected for GOLF during the red-wing mode by multiplying the amplitudes by 1.11.  For GONG, the resulting velocities were multiplied by 1.17 in order to take into account the different formation height of GONG; the factor was adjusted post-facto.  No other correction for formation height was applied to the remaining data sets.

We note two different amplitude regimes in the time series used by \citet{Fossat2017}, occurring at the beginning of the time series and after 2002 for about 2 years, both in the blue wing.  These two differences are related to the way that the temperature of the cathode of the photomultipliers is used to correct the calibration (Garcia, private communication, 2018).  We also note the variation of p-mode amplitude with the solar activity cycle with a typical variation of 9\% between solar maximum and solar minimum, as already measured by \citet{Howe2015} with the BiSON data.  Figure \ref{amplitude} indeed shows that a calibration that does not take into account the calibrated instrumental characteristics performs better in terms of a more consistent p-mode amplitude.  The most likely reason for this better performance may lie in the stability of the thermal environment provided after November 2002; thereby not requiring additional instrumental corrections.   Closer inspection of the p-mode amplitude also reveals that there is no yearly modulation present when GOLF observes in the blue wing.  On the other hand, a modulation is clearly seen when the observation is done in the red wing, which is due to the fact that the p-mode temperature fluctuations have a much larger contribution in that wing.

It is also interesting to compare the noise performance of the different instruments.  Again there is a clear difference for the GOLF noise between the two time series at the beginning of the time series (in the red wing) and after 2002 for about 2 years.  We also note that the GOLF noise does indeed increase with time primarily due to the photon noise increasing by a factor of five due to the number of photons dropping by a factor of 25.  Here, contrary to the p-mode amplitude, there is a clear yearly modulation of the noise in the GOLF data and in all data sets.

\section{Conclusion}
The advantage of the new calibration is that it requires only very limited knowledge of the instrument.  Although it may seem contradictory at first, the calibration is more in line with other instruments that are very well calibrated.  On the other hand, some aspects remain to be fully explained.  For instance, the yearly modulation in the velocity apparent in Fig.~\ref{residual}, which is possibly related to thermal effects induced by the variation of the SoHO--Sun distance throughout the orbit.  The current time series is also corrected for time shifts to keep { 99.8\%} of the corrected time series within $\pm$ 5 s.  There is a residual periodic time modulation due to the halo orbit, which is about 178 days.  In theory, the g-mode detection by \citet{Fossat2017} is not affected by either the changing amplitude of the p modes or the measured time shifts.  In practice, this new calibration approach may remove any doubts in the results of \citet{Fossat2017} related to these changes and time shifts.  This work will serve as a basis for an analysis of longer time series for testing the g-mode discoveries made by \citet{Fossat2017}.  

{ The new calibrated series are available as FITS files at www.ias.u-psud.fr/golf/assets/data/GOLF\_22y\_PM1, \textasciitilde/GOLF\_22y\_PM2 and \textasciitilde/GOLF\_22y\_mean for the PM1 and PM2 photomultipliers, and their mean, respectively.}




\begin{acknowledgements}
The GOLF instrument onboard SoHO is based on a consortium of institutes (IAS, CEA/Saclay, Nice and Bordeaux
Observatories from France, and IAC from Spain) involving a cooperative effort of scientists, engineers, and technicians, to whom we are indebted.  GOLF data are available at the MEDOC data and operations centre (CNES\,/\,CNRS\,/\,Univ. Paris-Sud) at medoc.ias.u-psud.fr.
SoHO is a mission of international collaboration between ESA and NASA.  This work uses data obtained by the Global Oscillation Network Group (\,GONG\,) program, managed by the National Solar Observatory, which is operated by AURA, Inc. under a cooperative agreement with the National Science Foundation. The data were acquired by instruments operated by the Big Bear Solar Observatory, High Altitude Observatory, Learmonth Solar Observatory, Udaipur Solar Observatory, Instituto de Astrof\'{\i}sica de Canarias, and Cerro Tololo Interamerican Observatory.  We thank Rafael Garc\'\i a, Bernhard Fleck, Antonio Jim\'enez, Ton von Overbeek, Catherine Renaud and William Thompson for fruitful discussions.
\end{acknowledgements}

\bibliographystyle{aa}
\bibliography{thierrya}

\begin{thebibliography}{16}
\expandafter\ifx\csname natexlab\endcsname\relax\def\natexlab#1{#1}\fi

\bibitem[{{Appourchaux} {et~al.}(2010){Appourchaux}, {Belkacem}, {Broomhall},
  {Chaplin}, {Gough}, {Houdek}, {Provost}, {Baudin}, {Boumier}, {Elsworth},
  {Garc{\'{\i}}a}, {Andersen}, {Finsterle}, {Fr{\"o}hlich}, {Gabriel}, {Grec},
  {Jim{\'e}nez}, {Kosovichev}, {Sekii}, {Toutain}, \&
  {Turck-Chi{\`e}ze}}]{Appourchaux2010}
{Appourchaux}, T., {Belkacem}, K., {Broomhall}, A.-M., {et~al.} 2010, \aapr,
  {\bf 18}, 197

\bibitem[{{Chaplin} {et~al.}(1996){Chaplin}, {Elsworth}, {Howe}, {Isaak},
  {McLeod}, {Miller}, {van der Raay}, {Wheeler}, \& {New}}]{WJC1996}
{Chaplin}, W.~J., {Elsworth}, Y., {Howe}, R., {et~al.} 1996, \solphys, {\bf
  168}, 1

\bibitem[{{Chidambararaj} \& {Sharma}(2016)}]{Halo2016}
{Chidambararaj}, P. \& {Sharma}, R.~K. 2016, International Journal of Astronomy
  and Astrophysics, {\bf 6}, 293

\bibitem[{{Elsworth} {et~al.}(1995){Elsworth}, {Howe}, {Isaak}, {McLeod},
  {Miller}, {New}, \& {Wheeler}}]{Elsworth1995}
{Elsworth}, Y., {Howe}, R., {Isaak}, G.~R., {et~al.} 1995, \aaps, {\bf 113},
  379

\bibitem[{{Feldman}(2011)}]{Feldman}
{Feldman}, M. 2011, {Hilbert transform applications in mechanical vibration}
  (John Wiley and Sons, Ltd)

\bibitem[{{Fossat} {et~al.}(2017){Fossat}, {Boumier}, {Corbard}, {Provost},
  {Salabert}, {Schmider}, {Gabriel}, {Grec}, {Renaud}, {Robillot},
  {Roca-Cort{\'e}s}, {Turck-Chi{\`e}ze}, {Ulrich}, \& {Lazrek}}]{Fossat2017}
{Fossat}, E., {Boumier}, P., {Corbard}, T., {et~al.} 2017, \aap, {\bf 604}, A40

\bibitem[{{Fr\"ohlich} {et~al.}(1997){Fr\"ohlich}, {Andersen}, {Appourchaux},
  {Berthomieu}, {Crommelynck}, {Domingo}, {Fichot}, {Finsterle}, {Gomez},
  {Gough}, {Jim\'enez}, {Leifsen}, {Lombaerts}, {Pap}, {Provost}, {Cort\'es},
  {Romero}, {Roth}, {Sekii}, {Telljohann}, {Toutain}, \& {Wehrli}}]{CF97}
{Fr\"ohlich}, C., {Andersen}, B.~N., {Appourchaux}, T., {et~al.} 1997,
  \solphys, {\bf 170}, 1

\bibitem[{{Gabriel} {et~al.}(1995){Gabriel}, {Grec}, {Charra}, {Robillot},
  {Roca Cort{\'e}s}, {Turck-Chi{\`e}ze}, {Bocchia}, {Boumier}, {Cantin},
  {Cesp{\'e}des}, {Cougrand}, {Cr{\'e}tolle}, {Dam{\'e}}, {Decaudin},
  {Delache}, {Denis}, {Duc}, {Dzitko}, {Fossat}, {Fourmond}, {Garc{\'{\i}}a},
  {Gough}, {Grivel}, {Herreros}, {Lagard{\`e}re}, {Moalic}, {Pall{\'e}},
  {P{\'e}trou}, {Sanchez}, {Ulrich}, \& {van der Raay}}]{Gabriel95}
{Gabriel}, A.~H., {Grec}, G., {Charra}, J., {et~al.} 1995, \solphys, {\bf 162},
  61

\bibitem[{{Garc{\'{\i}}a} {et~al.}(2005){Garc{\'{\i}}a}, {Turck-Chi{\`e}ze},
  {Boumier}, {Robillot}, {Bertello}, {Charra}, {Dzitko}, {Gabriel},
  {Jim{\'e}nez-Reyes}, {Pall{\'e}}, {Renaud}, {Roca Cort{\'e}s}, \&
  {Ulrich}}]{Garcia2005}
{Garc{\'{\i}}a}, R.~A., {Turck-Chi{\`e}ze}, S., {Boumier}, P., {et~al.} 2005,
  \aap, {\bf 442}, 385

\bibitem[{{Gelly} {et~al.}(2002){Gelly}, {Lazrek}, {Grec}, {Ayad}, {Schmider},
  {Renaud}, {Salabert}, \& {Fossat}}]{Gelly2002}
{Gelly}, B., {Lazrek}, M., {Grec}, G., {et~al.} 2002, \aap, {\bf 394}, 285

\bibitem[{{Harvey} {et~al.}(1996){Harvey}, {Hill}, {Hubbard}, {Kennedy},
  {Leibacher}, {Pintar}, {Gilman}, {Noyes}, {Title}, {Toomre}, {Ulrich},
  {Bhatnagar}, {Kennewell}, {Marquette}, {Patr{\'o}n}, {Sa{\'a}}, \&
  {Yasukawa}}]{Harvey1996}
{Harvey}, J.~W., {Hill}, F., {Hubbard}, R., {et~al.} 1996, Science, {\bf 272},
  1284

\bibitem[{{Howe} {et~al.}(2015){Howe}, {Davies}, {Chaplin}, {Elsworth}, \&
  {Hale}}]{Howe2015}
{Howe}, R., {Davies}, G.~R., {Chaplin}, W.~J., {Elsworth}, Y.~P., \& {Hale},
  S.~J. 2015, \mnras, {\bf 454}, 4120

\bibitem[{{Pall{\'e}} {et~al.}(1993){Pall{\'e}}, {Fossat}, {Regulo}, {Loudagh},
  {Schmider}, {Ehgamberdiev}, {Gelly}, {Grec}, {Khalikov}, {Lazrek}, \&
  {Sanchez}}]{Palle1993}
{Pall{\'e}}, P.~L., {Fossat}, E., {Regulo}, C., {et~al.} 1993, \aap, 280, 324

\bibitem[{{Renaud} {et~al.}(1999){Renaud}, {Grec}, {Boumier}, {Gabriel},
  {Robillot}, {Cort{\'e}s}, {Turck-Chi{\`e}ze}, \& {Ulrich}}]{Renaud1999}
{Renaud}, C., {Grec}, G., {Boumier}, P., {et~al.} 1999, \aap, 345, 1019

\bibitem[{{Schunker} {et~al.}(2018){Schunker}, {Schou}, {Gaulme}, \&
  {Gizon}}]{Schunker2018}
{Schunker}, H., {Schou}, J., {Gaulme}, P., \& {Gizon}, L. 2018, \solphys, 293,
  95

\bibitem[{{Ulrich} {et~al.}(2000){Ulrich}, {Garc{\'{\i}}a}, {Robillot},
  {Turck-Chi{\`e}ze}, {Bertello}, {Charra}, {Dzitko}, {Gabriel}, \& {Roca
  Cort{\'e}s}}]{Ulrich2000}
{Ulrich}, R.~K., {Garc{\'{\i}}a}, R.~A., {Robillot}, J.-M., {et~al.} 2000,
  \aap, {\bf 364}, 799

\end{thebibliography}

\appendix
\section{Timing of helioseismic instruments}
The time of the mid point of the first sample for each helioseismic instruments is given in Table~\ref{timing}.  For GONG, because of the use of the BDF, the first sample $x$ is computed  by taking $x_{\rm GONG}(t_0)={\rm v}(t_0)-{\rm v}(t_0-1)$ where $t_0$ is the time in minutes given in the file (for GONG the file directly provides $x_{\rm GONG}(t_0)$) and ${\rm v}(t_0)$ is the velocity.  Because of the use of the BDF for GONG, the time delay is computed by also taking the BDF of the other instruments, that is by computing $x_{\rm Other}(t_1)={\rm v}(t_1+1)-{\rm v}(t_1)$, where $t_1$ is the time in minutes given in the file. For the other instruments, the file directly provides ${\rm v}(t_1)$; therefore when $t_0=t_1$, the time difference between GONG and the other instruments is -60 s.

\begin{table*}[!]
\begin{tabular}{c c c c c c c c} 
\hline
\hline
        Instrument &  Cadence &  Mid point &  Mid point &  Mid point & Raw time difference  & Time corrected  & Time measured \\
        & (s) & UT & GPS & TAI &w.r.t. GOLF (s) &when use of BDF (s) & with p modes \\
\hline
\hline
GOLF$^{1}$ &    60      & 00:00:30.0 &  00:00:41.0 &    00:01:00.0      & 0       & 0     &0         \\
GONG$^{2}$ &    60  &   00:00:35.0 &    00:00:46.0      & 00:01:05.0    & +5 &    -55     & -48$^{4}$ \\
MDI$^{3}$ &     60      & 23:59:30.0 &  23:59:41.0 &    00:00:00.0 &    -60 &       -60 &   -48        \\
BiSON$^{1}$&    40 &    00:00:20.0 &    00:00:31.0 &    00:00:50.0 &    -10 &       -10 &   2.8  \\ 
VIRGO$^{3}$&    60 &    23:59:34.8 &    23:59:45.8 &    00:00:04.8 &    -55.2 &       -55.2   & 25     \\
\hline
\hline
\end{tabular}
\\
$^{1}$ Time given in the file corresponds to start of integration\\
$^{2}$ Time given in the file corresponds to start of integration, and does not take into account the BDF time "effect"\\
$^{3}$ Time given in the file corresponds to mid point\\
$^{4}$ Comparison with GONG is done by applying the BDF to GOLF as well
\caption{Time difference compared to GOLF measured using p modes compared to expected time difference for May 1 1996.  First column gives the instrument, the second column the cadence of the instrument, the third to fifth columns the time of the first sample in Universal Time (UT), from the Global Positioning System (GPS) and TAI references; the sixth column the time difference with respect to (w.r.t.) GOLF; the seventh column gives the time corrected when comparing to GONG since GONG consortium delivers BDF filtered data ; the last column gives the time difference measured with p modes.}

\label{timing}
\end{table*}

\end{document}